# Damage accumulation during high temperature fatigue of Ti/SiC$_f$ metal matrix composites under different stress amplitudes

Ying Wang[1,2], Xu Xu[1], Wenxia Zhao[3], Nan Li[3], Samuel A. McDonald[2,†], Yuan Chai[2], Michael Atkinson[1], Katherine J. Dobson[4], Stefan Michalik[5], Yingwei Fan[3], Philip J. Withers[1,2], Xiaorong Zhou[1] and Timothy L. Burnett[1,2,*]

[1] Department of Materials, The University of Manchester, Manchester, M13 9PL, UK

[2] Henry Royce Institute for Advanced Materials, Department of Materials, The University of Manchester, Manchester, M13 9PL, UK

[3] Beijing Institute for Aeronautical Materials, Aero Engine Corporation of China, Beijing, 100095, China

[4] Department of Civil and Environmental Engineering, University of Strathclyde, Glasgow, G1 1XQ, UK

[5] Diamond Light Source Ltd., Harwell Science and Innovation Campus, Didcot, OX11 0DE, UK

[†]Present address: MAX IV Laboratory, Lund University, PO Box 118, 221 00 Lund, Sweden

*Corresponding author. E-mail address: timothy.burnett@manchester.ac.uk

**Abstract**

The damage mechanisms and load redistribution of high strength TC17 titanium alloy/unidirectional SiC fibre composite (fibre diameter = 100 μm) under high temperature (350 °C) fatigue cycling have been investigated *in situ* using synchrotron X-ray computed tomography (CT) and X-ray diffraction (XRD) for high cycle fatigue (HCF) under different stress amplitudes. The three-dimensional morphology of the crack and fibre fractures has been mapped by CT. During stable growth, matrix cracking dominates with the crack deflecting (by 50-100 μm in height) when bypassing bridging fibres. A small number of bridging fibres have fractured close to the matrix crack plane especially under relatively high stress amplitude cycling. Loading to the peak stress led to rapid crack growth accompanied by a burst of fibre fractures. Many of the fibre fractures occurred 50-300 μm from the matrix crack plane during rapid growth, in contrast to that in the stable growth stage, leading to extensive fibre pull-out on the fracture surface. The changes in fibre loading, interfacial stress, and the extent of fibre-matrix debonding in the vicinity of the crack have been mapped for the fatigue cycle and after the rapid growth by high spatial resolution XRD. The fibre/matrix interfacial sliding extends up to 600 µm (in the stable growth zone) or 700 µm (in the rapid growth zone) either side of the crack plane. The direction of interfacial shear stress reverses with the loading cycle, with the maximum frictional sliding stress reaching ~55 MPa in both the stable growth and rapid growth regimes.

# 1   Introduction

Unidirectional SiC fibre reinforced Ti alloy metal matrix composites (MMCs) are attractive candidate materials for use in bladed discs (BLISCS) and rings (BLINGS) in the compressor stage of aero-engines, due to their high specific stiffness, high specific strength, and capability to operate at high service temperatures (~350 °C) [1]. Use of these materials could contribute to ~40 % weight reduction over unreinforced components [2,3], providing significant benefits to aero-engine design. The fatigue performance of Ti alloy/SiC fibre composites is a crucial aspect in such applications, especially at elevated temperatures. The fibre reinforcement has been found to improve the fatigue crack growth resistance in alloys at room temperature (RT), e.g. [4]; while the fatigue life at elevated temperatures that would be experienced in operando has received less attention. At RT, fibre bridging is an important mechanism shielding the fatigue crack and lowering the effective stress intensity factor at the crack tip leading ultimately to crack arrest, if the bridging fibres do not fracture [5]. At elevated temperatures (200-500 °C), Cotterill and Bowen [6] reported sharp jumps in the crack growth (by up to two orders of magnitude) associated with discrete bridging fibre fractures during fatigue crack propagation in Ti-15-3/SCS6 composites, contributing to shorter fatigue lives compared with at RT.

It is known that the interfacial sliding/shear strength (ISS) is a key controlling parameter as to whether the matrix crack bypasses or fracture the fibres, as the ISS determines the level of load transfer to the fibres once debonding occurs [7,8]. During manufacturing, the cooling process causes radial clamping stressed on the fibres exerted by the matrix, due to the mismatch in coefficients of thermal expansion. This radial clamping stress decreases with increasing temperature, thus reducing Coulombic friction and leading to a lower interfacial sliding stress at elevated temperatures [9]. In addition, the nature of the interface can change significantly on exposure to high temperatures. Interfacial oxidation can occur at temperatures above 300 °C [10], which embrittles the fibre/matrix interface exposed along the bridged crack and promotes crack development. Moreover, the degradation in fibre strength produced by the combined effect of elevated temperature and the environment may explain the tendency for increased fibre fracture over fibre bridging and crack arrest at elevated temperature [6,11]. However, very limited information is available concerning the different crack growth behaviours and stress redistribution mechanisms associated with fibre bridging and fibre fracture in Ti alloy/SiC fibre composite at temperatures representative of service conditions within aeroengines.

Use of time-lapse X-ray computed tomography (CT) has been growing to assess damage evolution in materials owing to its three-dimensional (3D) non-destructive imaging capability [12,13]. It is particularly well suited to the study of multi-phase or complex-structured materials such as composites [14,15]. Alongside this, X-ray diffraction (XRD) is able to provide important information on the partitioning of load between reinforcement and matrix of Ti/SiC composite materials [16]. Hard synchrotron X-ray diffraction can be used to map the strain distribution inside the material and crucially between the phases with high spatial resolution [17]. The two methods have been used in concert to study the damage development and local load partitioning between fibres and matrix during fatigue crack bridging in Ti–6Al–4V/SCS-6 (carbon cored) SiC fibre composites at RT [7,18,19]. In addition, the load partitioning mechanisms of Ti alloy/SiC fibre composites during fatigue crack bridging at elevated temperature has also been examined in [20]. However, the 3D damage morphology and the load redistribution associated with sudden bursts in fibre fractures have not been reported to date.

The aim of this paper is to investigate the high-temperature damage accumulation mechanisms and load partitioning in high strength TC17 titanium alloy/SiC fibre composites. Firstly the behaviour during stable fatigue crack growth at low fatigue stress amplitude is studied, where fibres bridge the fatigue crack as it propagates. This is compared with the behaviour at the peak stress in the higher stress amplitude fatigue case, where rapid crack growth and a burst of fibre fractures occurred. Here the 3D damage morphology was characterised by X-ray CT, while the fibre strain distribution and interfacial stress are mapped by high spatial resolution XRD. The observed differences in damage accumulation behaviours between the two crack-development scenarios provide insights into the high-temperature damage mechanisms in TC17/SiC fibre composites in operando.

## 2    Materials and Methods

### 2.1    *Sample preparation*

Here the behaviour of unidirectional SiC fibre reinforced TC17 titanium alloy (Ti-5Al-2Sn-2Zr-4Mo-4Cr) composite [21] is studied. The SiC fibres are 100 μm in diameter, having a 15 μm diameter W core and a 2-3 μm outer carbon coating. In brief, the composite was manufactured by the consolidation of TC17 alloy coated SiC fibres (4 × 12 array of close packed arrangement) encapsulated in a pre-machined TC17 can at 920 °C giving rise to samples measuring 120 × 4 × 2 mm (see Fig. 1(a)). In the reinforced region the fibre volume fraction was ~45%. Electron discharge machining (EDM) was used to remove some of the TC17 cladding from the sample surfaces and to machine the test-piece geometries, resulting in a 20 mm long gauge section with 1.2 mm thick by 4 mm wide cross-section (see Fig. 1(a)). Some cladding (~330 µm) was retained on each side. A through-thickness edge notch (~1 mm deep) was introduced into each sample by EDM to act as a crack initiator. However, in preliminary tests it was found that multiple cracks tended to initiate from the relatively blunt tip of the EDM notch as has been observed previously for similar systems [19]. Therefore, in order to initiate a single fatigue crack, a sharper notch (~0.05 mm deep) was introduced by high pulse laser ahead of the EDM notch using a microPREP™ milling machine (3D-Micromac), see Fig. 1(b).

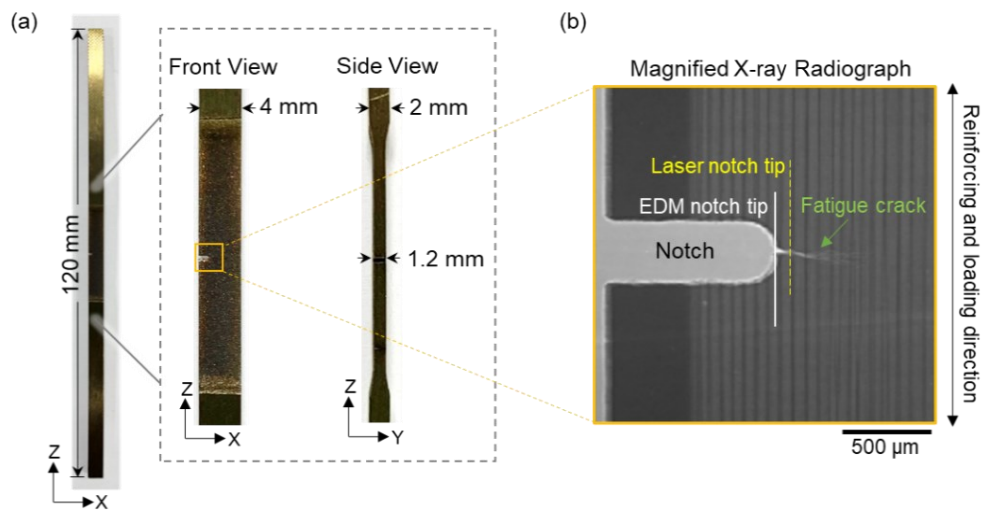

Fig. 1. Geometry of the test specimen (S1). (a) Photographs of the front and side views showing the specimen dimensions. (b) X-ray radiograph of the region around the notch as marked on the magnified front view in (a), showing the notch introduced by EDM and laser machining, and the fatigue crack.

The samples were pre-fatigued on an Instron servo-hydraulic testing machine (equipped with an environmental chamber) at 350 °C under tension-tension loading at a stress ratio $R$ (where $R = K_{min}/K_{max}$) of 0.1 applied at a frequency of 10 Hz. $K_{min}$ and $K_{max}$ correspond to the minimum and maximum stress-intensity factors applied within a fatigue cycle, and the initial stress intensity factor range $\Delta K_{ini} = K_{max} - K_{min}$. Both low (sample S1) and higher (sample S2) fatigue stress ranges were studied where $\Delta K_{ini}$ was 14.9 MPa√m ($K_{max}$ = 16.5 MPa√m) for S1 and 19.0 MPa√m for S2 ($K_{max}$ = 21.1 MPa√m). Fig. 2(a) illustrates the fatigue testing scenario and Table 1 summarises the fatigue conditions of the two specimens for synchrotron experiment. The fatigue test was interrupted at 10,000 cycles and the samples were imaged by X-ray radiography on a Nikon XTH 225 CT system to assess the crack length. The crack lengths (including the notch) were measured to be around 1.6 and 2.0 mm for S1 and S2, respectively (see Table 1).

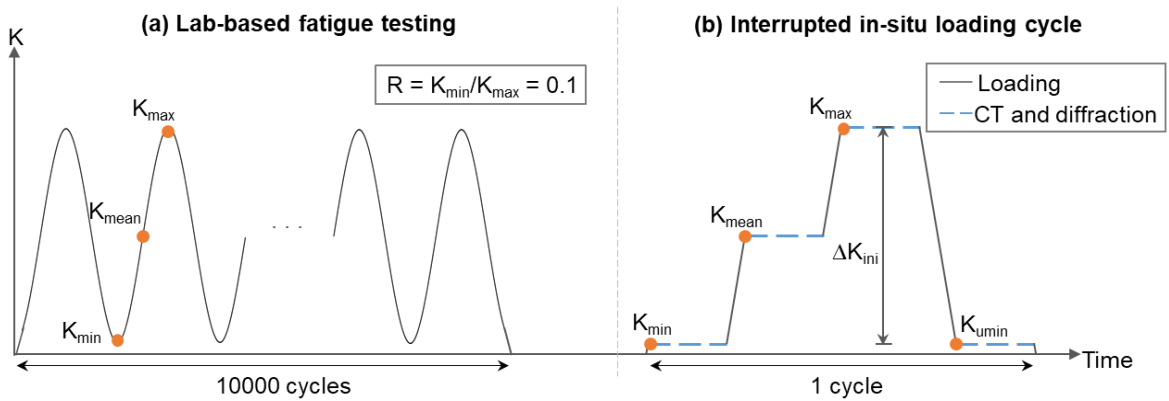

Fig. 2. The loading history of the tested specimens. The specimens were (a) pre-fatigued to initiate and grow fatigue cracks, followed by (b) *in situ* loading on the synchrotron beamline to image the damage and map the stress distribution local to the crack as a function of load through the fatigue cycle.

Table 1. Summary of prior fatigue testing for the low (S1) and higher (S2) stress amplitude test samples.

| Specimen | $\Delta K_{ini}$ (MPa√m) | Max Fatigue Load (N) | No. of Cycles | Notch Depth (mm) | Crack Length (mm) |
|---|---|---|---|---|---|
| S1 | 14.9 | 865 | 10000 | 1.0 | 1.6 |
| S2 | 19.0 | 1200 | 10000 | 1.0 | 2.0 |

## 2.2  *In situ synchrotron X-ray tomography and high resolution X-ray diffraction*

The *in situ* experiment was performed on the I12 beamline [22] at Diamond Light Source (DLS), Oxfordshire, UK. A monochromatic X-ray energy of 53 keV (wavelength = 0.233Å) was selected for both X-ray CT and XRD. All measurements were carried out in transmission. A PCO.edge camera was employed to acquire the X-ray CT datasets to visualise the fatigue crack path using optical module 3 (pixel size 3.25 µm, field of view 7 mm × 8 mm ($x \times z$)). A 2M CdTe Pilatus 2D area detector was employed to acquire the Debye Scherrer diffraction rings. The area (marked in dashed yellow box on the radiograph in Fig. 3(b)) encompassing the fatigue crack was mapped by XRD using a beam size (determined by the incident slit size) of 150 × 50 µm ($x \times z$). The step size was 25 µm for $|z| \leq 1$ mm and 50 µm for $|z| > 1$ mm (with the crack located at $z = 0$), in order to obtain higher resolution closer to the fatigue crack. The exposure time was 2 seconds per measurement point.

The Deben-Manchester Open Frame Rig was used for *in situ* loading and heating [23]. It was fitted with a bespoke high-temperature furnace [24], and was mounted on the sample stage on the I12 beamline (see Fig. 3(a)). The fatigued specimens (S1 and S2) were heated to 350 °C while the tensile load was actively maintained at 0 N, then quasi-statically loaded under tension through a fatigue cycle on the beamline following the scheme shown in Fig. 2(b). By translating the imaging and diffraction detectors into and out of the beam, both tomography and diffraction measurements were acquired at $K_{min}$, $K_{mean}$ and $K_{max}$ during loading and $K_{umin}$ during unloading for each specimen (see Fig. 2(b)).

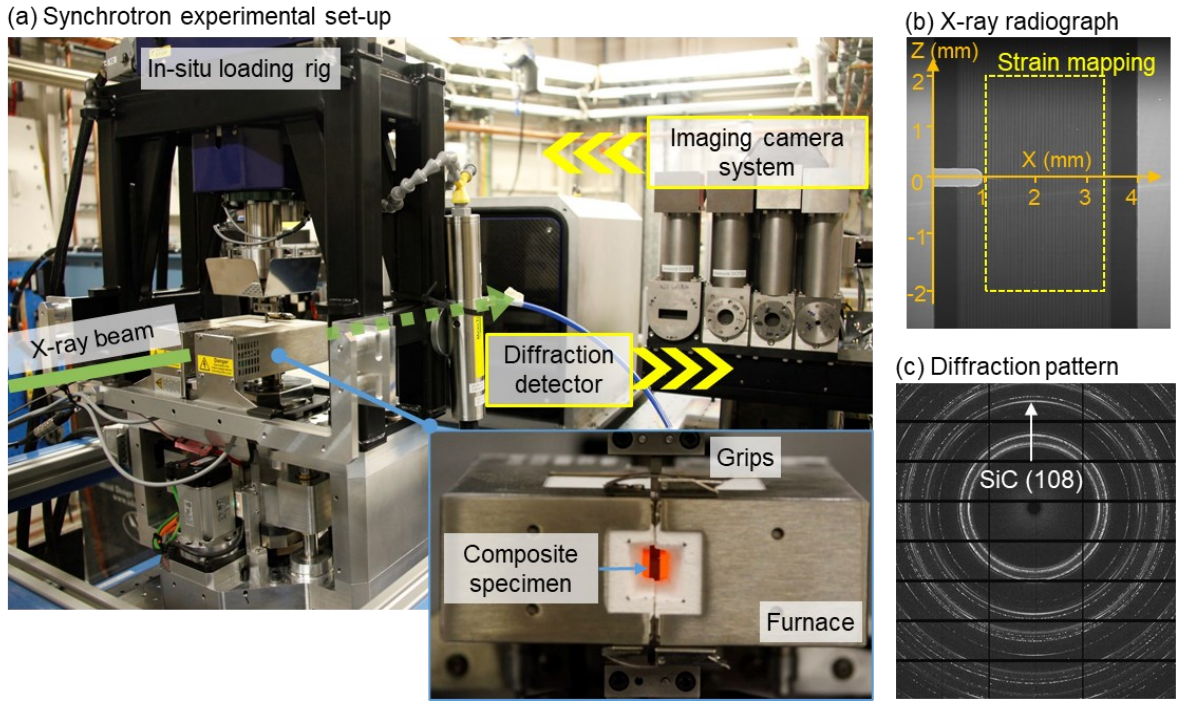

Fig. 3. Experimental set-up for *in situ* transmission XRD and X-ray CT on the I12 beamline at Diamond Light Source. (a) Photograph showing the *in situ* loading rig. The inset shows the gauge length of the sample within the furnace mounted on the *in situ* loading rig, while the grips are located outside of the furnace. (b) A typical X-ray radiograph, on which the X and Z coordinate system is indicated. (c) A typical diffraction pattern indicating the (108) SiC reflection used to infer the fibre strains.

For the low fatigue stress amplitude sample (S1), the crack grows stably, whereas for the higher fatigue stress amplitude test sample (S2), a period of rapid crack growth was observed accompanied by a 100 N load drop upon loading the sample to $K_{max}$. The specimen was then reloaded to the maximum fatigue load, and no further crack growth occurred. It should be noted that for specimen S1, a hardware issue meant that the data at $K_{min}$ was not successfully recorded.

*2.3    Strain measurement*

In order to improve the counting statistics, the 360° diffraction rings were binned azimuthally into twelve 30° wide segments. The 'caked' segments centred around the north and south poles were summed azimuthally (over an angular range of ± 15°) to provide line profiles of diffracted intensity versus $2\theta$ using DAWN software (DLS) [25]. The (108) SiC reflection (see Fig. 3(c)) was used to infer the fibre strains. The reflections for the titanium phases were not quantified, because the transmitted signal is averaged over the whole path length through the samples which includes both the reinforced and un-reinforced Ti-alloy regions, making it unrepresentative of the alloy behaviour within the composite. It should be noted that each specimen contained roughly four plies of SiC fibres, thus each measurement point represents an average strain over all fibres through the thickness having the same X and Z positions. The centre ($2\theta$) of the (108) SiC peak was found by Pseudo-Voigt peak fitting using Python. By applying Bragg's Law, the $2\theta$ data was converted into $d$ spacing *values*. The axial elastic strain, $\varepsilon_f$, in the fibre is then inferred from the change in lattice spacing;

$$\varepsilon_f = \frac{d - d_0}{d_0} \qquad (1)$$

where $d_0$ corresponds to the "stress-free" $d$-spacing. Here, $d_0$ was not directly measured on exposed fibre, but estimated assuming that at the end of a broken SiC fibre in the composite the axial stress equals zero following Hung et al. [20]. Moreover, it is well established that the interfacial shear stress (ISS), $\tau$, can be inferred using stress balance from the gradient in the fibre strain, $\varepsilon_f$, along the fibre [26]. The ISS averaged over all the fibres along the X-ray path length can therefore be obtained using the following equation

$$\tau = \frac{E_f r_f}{2} \times \frac{d\varepsilon_f}{dz} \qquad (2)$$

where $E_f$ is the fibre Young's modulus, $r_f$ is the fibre radius and $d\varepsilon_f/dz$ represents the fibre strain gradient along the fibre.

## 2.4 X-ray CT

For each CT scan, 1200 X-ray radiographs/projections (exposure time 0.3 s per projection) were acquired over 180° rotation. The 2D projections were reconstructed into 3D CT volumes using the SAVU reconstruction pipeline [27] developed at DLS, which employs a standard filtered back-projection algorithm. The reconstructed CT data was imported into Avizo 2019.1 (Thermo Fisher Scientific) and MATLAB R2017 (Mathworks) for further image analysis and quantification. The 3D crack was segmented in Avizo using the following steps, 1) extraction of the composite volume as a region-of-interest by thresholding to remove the background; 2) use of the top hat segmentation tool in XZ sections of the composite volume with line seeds lying along the X direction to extract the crack; and 3) use of the 3D grow, fill and shrink tools to refine the segmentation. The segmented fatigue crack was then rendered in 3D for visualisation.

*2.5   Fractography*

After X-ray imaging and strain mapping, the sample S2 was removed and placed in liquid nitrogen before fast fracture in monotonic tension. This facilitated a clear visualisation of the fatigue and fast-fracture regions on the fracture surface. The fracture surface was examined using a Tescan MIRA3 Field Emission Gun Scanning Electron Microscope (FEG-SEM) collecting the secondary electron signal at an accelerating voltage of 5 kV.

**3    Results and Discussion**

*3.1   Behaviour in the stable growth regime*

Under low stress amplitude fatigue (sample S1) stable crack growth was observed and in this section we look at the nature of the fatigue crack and the load partitioning between matrix and fibres in the vicinity of the crack.

*3.1.1   Fatigue crack morphology and fibre bridging in 3D*

The 3D fatigue crack morphology during stable crack growth can be appreciated from the CT scan at $K_{max}$ for S1. It is evident from the magnified view of the XZ section in Fig. 4 that, in contrast to Barney et al. [28] and Hung et al. [19], a single fatigue crack initiated from the notch, promoted by sharpening the notch tip by laser machining. The presence of the Ti cladding is clear in the XY view.

The fatigue crack within the fibre reinforced composite region, as highlighted by the blue box on the XY section in Fig. 4, was segmented and extracted for 3D visualisation in Fig. 5. It can be seen that the fatigue crack grows through the matrix bypassing the fibres which remain intact in agreement with previous work [5,19,29]. This imparts significant crack bridging and hence crack-tip shielding during stable fatigue crack growth. There is no evidence from the CT of the propagation of any opened cracks along the interfaces. The 3D volume rendering (see Fig. 5) shows that a single crack is propagating but is deflected along its path. Although slightly shorter on one side than the other (probably due to there being more cladding on one side than the other), the fatigue crack front is relatively straight. There is no evidence of the middle of the crack front being significantly behind or ahead of the edges of the crack.

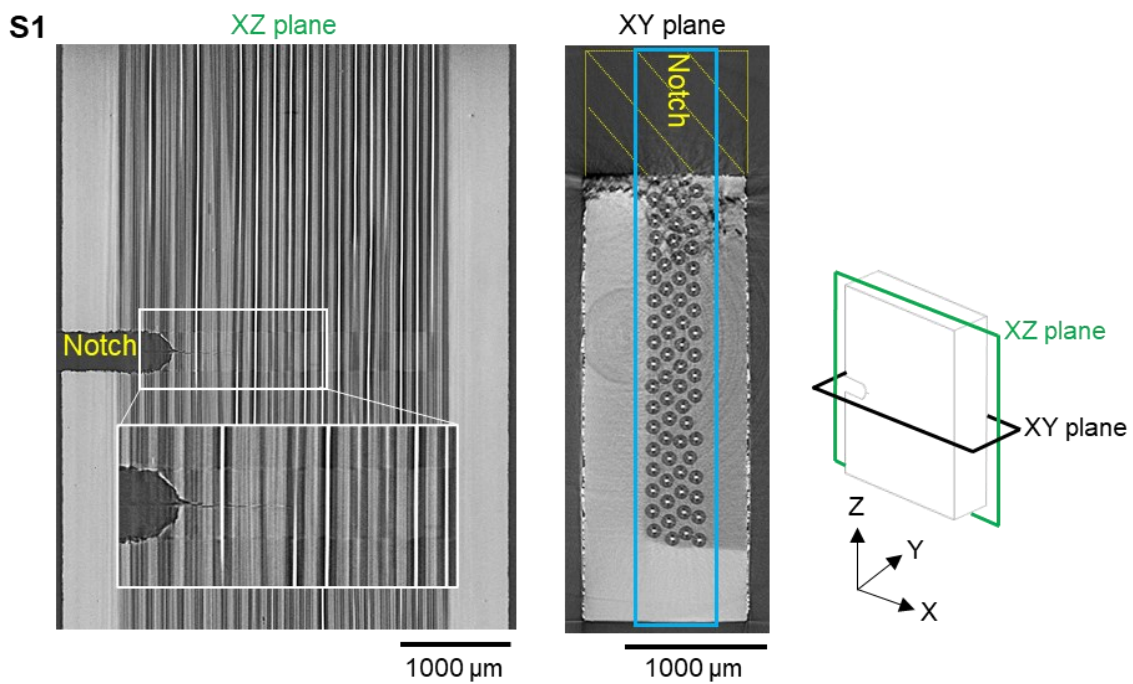

Fig. 4. Fatigue crack morphology during the stable-growth stage. X-ray CT virtual XZ and XY sections for S1 at $K_{max}$ (16.5 MPa√m). The fibre reinforced region, which will be analysed further, is highlighted by the blue box on XY sections. The fact that the crack surface is not completely flat means that it is difficult to discern in the XY section. The W cores are evident as bright contrast, the Ti matrix as light grey the SiC fibre somewhat darker and the cracks darkest of all.

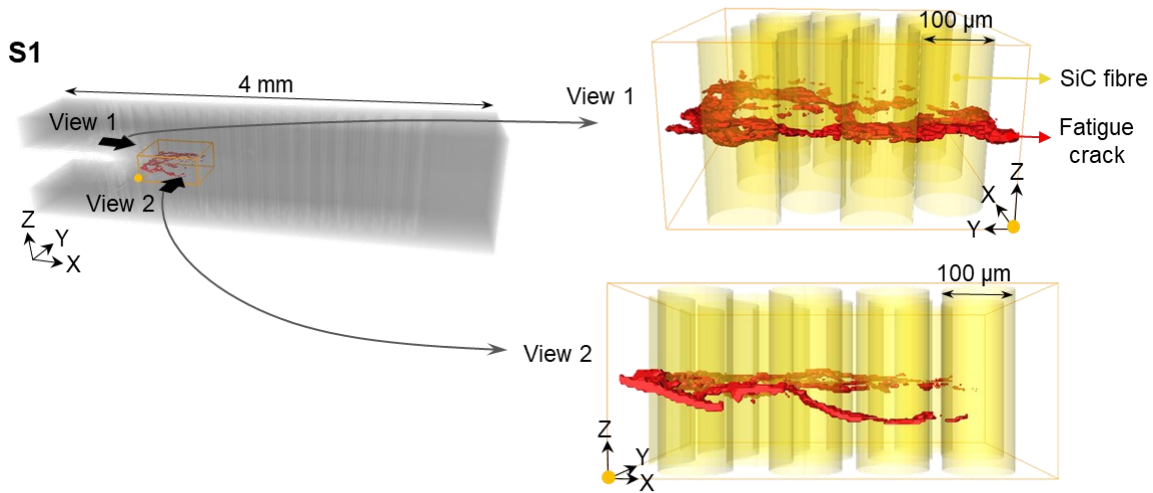

Fig. 5. 3D fatigue crack morphology during the stable-growth stage, showing fibre bridging and crack deflection. 3D volume rendered X-ray CT image of the fibre reinforced region of interest (grey), with the fatigue crack (rendered red) and SiC fibres (rendered yellow) for S1 (low $\Delta K_{ini}$) at $K_{max}$ (16.5 MPa√m). The magnified views (with the matrix rendered transparent) show head on (View 1) and side on (View 2) views of the crack.

Based on the 3D crack renderings in Fig. 5, the deviation of the fatigue crack from the median crack plane was quantified in MATLAB and shown in Fig. 6. This shows considerable crack deflection as the crack bifurcates to run around each fibre, in common with the observations of Hung *et al.* [19]. The extent of the crack deviation tends to be less than 100 μm from the median plane.

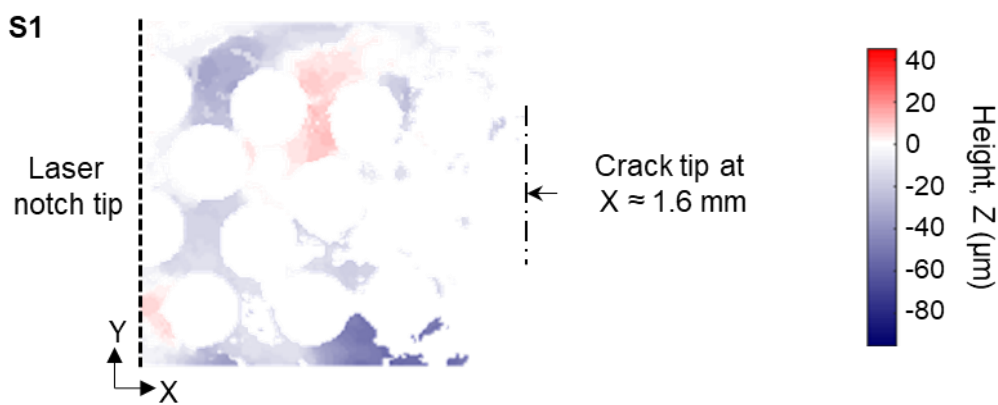

Fig. 6. The map of crack height (distance from the median plane) for S1 at $K_{max}$ (16.5 MPa√m) showing fatigue crack deflection during the stable-growth stage. The height of the fatigue crack plane above and below z = 0 mm (the height of notch tip) is calculated based on the X-ray CT segmentation of the crack.

*3.1.2 Micro-mechanisms of stress redistribution between matrix and fibres*

The strain distribution in the SiC fibres local to the crack has been mapped by X-ray diffraction as described in section 2.3. Fig. 7(a) shows the relative axial elastic strain maps for the fibres in S1 at $K_{max}$ and $K_{umin}$ in the region of the crack tip (located at x = 1.6 mm). It should be remembered that each measurement is collected in transmission through the entire specimen and is an average for all fibres (along the Y axis) at each measurement position, (X, Z). This results in some convolution effects if individual fibres bear different loads. Given that the fatigue crack front was relatively straight and flat, the stress conditions should be similar through the specimen thickness.

It is well documented that thermal residual stresses are introduced into the Ti/SiC composites during cooling from the processing temperature, owing to the difference in the thermal expansion coefficient between SiC fibres and Ti-alloy matrix [16]. At room temperature these stresses are compressive in the fibres, both axially and radially, and are somewhat relieved by heating to 350 °C. Here the axial compressive strain in the fibres located far from the crack in the unloaded state is ~ -0.13% at 350 °C, which is slightly less than observed previously at 300 °C (-0.16%) and 120 °C (-0.19%) for Ti-6Al-4V/SCS-6 composites [20], as one might expect given the higher testing temperature. Wedge-shaped stress concentration zones are observed in the fibre strain maps in Fig. 7(a); these were also seen in fatigue cracked Ti-4Al-6V/SCS-6 composites [7,20] and are characteristic of a local redistribution of the tensile load to the fibres in the vicinity of the crack.

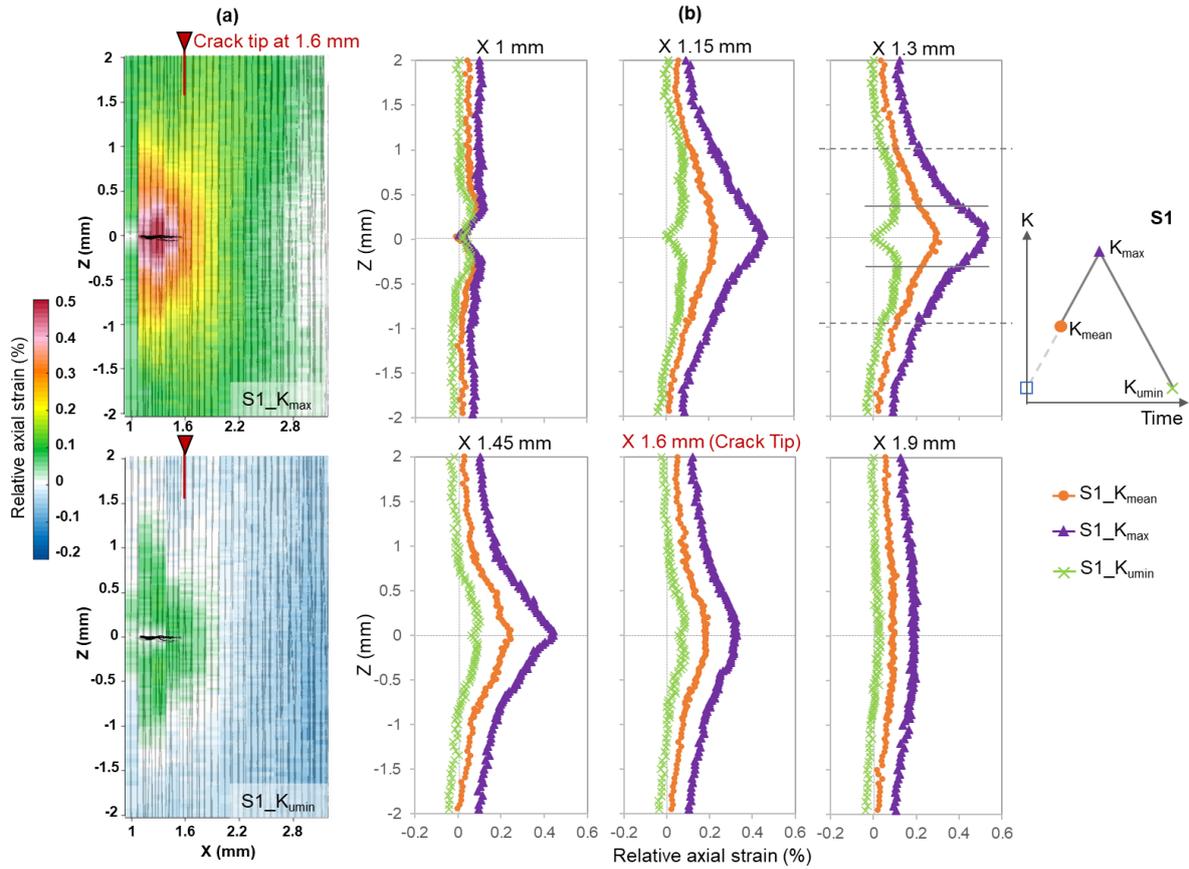

Fig. 7. Variation in relative axial elastic strain along the fibres as a function of load and position (where z = 0 is the notch-tip height) for S1 at 350 °C. (a) The fibre strain map at $K_{max}$ (top) and $K_{umin}$ (bottom); superimposed on a rendering of the W cores of the SiC fibres (in grey vertical lines) and the ~1.6 mm fatigue crack (rendered black). (b) The fibre elastic strain profiles at X positions of 1, 1.15, 1.3, 1.45 (in the crack wake), 1.6 (at the crack tip) and 1.9 mm (ahead of crack tip) recorded at $K_{mean}$, $K_{max}$ and $K_{umin}$. In a) each coloured pixel corresponds to one through-thickness averaged diffraction measurement, i.e. over a window of 150 μm (along the X axis) × 50 μm (along the Z axis).

The strain/stress distribution along the SiC fibres both behind and ahead of the fatigue crack tip are best viewed as line profiles to investigate the effect of fatigue crack on load redistribution over a fatigue cycle during stable crack growth as displayed in Fig. 7(b) for S1. At x = 1 mm (see Fig. 7(b)), the axial fibre strain at the median plane (z = 0) is approximately zero at all load levels, which indicates that the fibres at the notch were damaged by the laser notching process. As illustrated by the strain profiles at x = 1.3 mm (see Fig. 7(b)), in regions outside the dotted horizontal lines far from the fatigue crack plane (|z| > 0.8 mm), the response in fibre strain to loading and unloading is approximately proportional to the applied

load at each height. This indicates that these regions of the interface were intact and were free from interfacial sliding during the cyclic loading. In regions closer to the crack plane ($|z| <$ 0.8 mm), the change in fibre strain in response to applied load is no longer proportional, which indicates the presence of interfacial debonding and sliding. The debonding length is rather similar to previous observations in Ti-6Al-4V/SCS-6 composites fatigued under three-point bending at room [7] and elevated [20] temperatures. It is noteworthy that in Fig. 7(b) the slope in the fibre strain profile changes either side of the horizontal solid lines; this represents a change in the direction of the interfacial stress (Equation (2)), where the kink at $|z| \approx 0.3$ mm corresponds to the transition point. The fact that the load redistribution behaviour in the region between the dotted and solid lines is different between those regions adjacent to the crack plane could indicate a partial-sliding/sticking region [9], or this could be due to the effect of the prolonged exposure to high temperature on the local interfacial properties [30] during the *in situ* measurement, or could be due to the convolution of the different conditions experienced by different fibres through the thickness. For fibres positioned well ahead of the fatigue crack tip (e.g. x = 1.9 mm in Fig. 7(b)) the variation in strain along the fibre is much less evident. During unloading, reverse-sliding occurred in the debonded regions, and the 'm' shape in the unloaded profile (green) is evidence of some frictional resistance inhibiting the reverse-sliding of the fibres bridging the crack [31]. Moreover, the fibre bridging stress across the fatigue crack wake can be estimated based on the fibre strains around z = 0, which marks the approximate crack plane. The maximum fibre bridging stress was calculated to be ~ 2100 MPa (at an axial strain of ~ 0.52%) across the crack in S1, which is about ~ 1.5 times of that reported previously for Ti-6Al-4V/SCS-6 composite at 300 °C [20]. The fact that the fibres bridging the crack carry part of the applied load acts to lower the effective crack-tip driving force during the stage of stable fatigue crack growth [5,32].

*3.1.3 Interfacial sliding during stable growth at elevated temperature*

As discussed in Section 2.3, the variation in fibre strains along the fibre direction and with distance from the crack tip in Fig. 7 reflects different interfacial and stress transfer conditions. When using Equation (2) to infer the ISS from the strain variation along the fibres it should be remembered that the position of the crack plane varies slightly across the specimen. This causes a smearing of the ISS variation due to a contribution from all the fibres throughout the thickness. Fig. 8 shows the ISS along fibres at different positions with respect to the crack tip for S1 calculated from the corresponding fibre strain profiles, where the sign of the ISS indicates the direction of the shear stress. The curves are broadly similar to those described for Ti-6Al-4V/SCS6 system [19] and the reader is referred to the schematics in that paper for more details as to their interpretation. In essence, the change of sliding direction (forward sliding under tensile loading as the fibres are pulled out of the matrix slightly and reverse sliding during unloading when the fibres are pushed back in) is evident in terms of a change in sign of the ISS between $K_{max}$ and $K_{umin}$. Sliding occurs when the shear stress reaches the interfacial frictional sliding stress ($\tau_{fr}$) causing the shear stress to plateau. At x = 1.3 mm (0.3 mm behind the crack tip) this occurs at approximately $\tau_{fr}$ = 55 MPa (at $K_{max}$) reversing in sign at $K_{umin}$ suggesting sliding lengths (denoted by the boxes in Fig. 8) of ~400 μm in the forward direction and just ~100 μm in the reverse direction. The curves are somewhat noisy, but at x = 1.15 mm the frictional sliding stress plateau is lower ($\tau_{fr} \approx$ 35 MPa) and the sliding zone extends a little further from the crack plane (~600 μm). For x = 1 mm (see Fig. 8), there is very little change in the ISS profile between $K_{max}$ and $K_{umin}$, suggesting that there is no sliding and the shear stress introduced when the fibres at the notch fractured is largely retained through the whole fatigue cycle. Immediately behind the crack tip (x = 1.45 mm) there is evidence of forward sliding, but reverse sliding is difficult to discern. This may be because the sliding lengths are even smaller than those for x = 1.3 mm and so difficult to measure or because the fibres further behind the crack hold the crack open such that the fibres very close to the crack slide back only a very short distance into the matrix. At the crack tip (x = 1.6 mm) and ahead of the crack tip (x = 1.9 mm) no forward and reverse frictional sliding is detected. That the ISS is not constant within the sliding region was also observed by Withers et al. [9] in single-fibre-fragmentation testing. This was also observed in Ti-6Al-4V/SCS-6 composite fatigued at 300 °C, but not seen in composites fatigued at RT or 120 °C [7,20] and may be due to damage in the carbon coating. The decrease in frictional sliding stress for fibres located towards the crack mouth may be because of damage introduced to the interface by the many fatigue cycles as the crack propagates as discussed in Hung et al. [19].

The effect of temperature on frictional sliding stress has previously been investigated for pristine interfaces [9]. In that case the frictional ISS decreased from ~150 MPa at RT to ~50 MPa at 400 °C in Ti-6Al-4V/SCS-6 (C core fibres) composite, and from 160-170 MPa at RT to 60-100 MPa at 400 °C in Ti-6Al-4V/SM2156 (W core fibres) composite. For worn (fatigued) interfaces, the ISS was reported to decrease slightly with increasing temperature, ~60 MPa at RT, 40-50 MPa at 120 °C and 20-40 MPa at 300 °C in fatigued Ti-6Al-4V/SCS-6 (C core) composite [20]. The reduction in frictional ISS at elevated temperature comes partly from the asperity wear of the coating layer after debonding and partly from the relaxation of radial thermal stress in the matrix which clamp the fibres [33]. Here, the maximum forward sliding stress observed is ~55 MPa during loading at 350 °C, indicating that the TC17/SiC fibre (W core) composite seems to have a slightly stronger interface compared with the previously reported Ti/SiC SCS-6 composites at elevated temperature. Upon unloading, the maximum ISS for reverse sliding is close to the maximum ISS in the forward direction for S1. This is in contrast to the observations in both pristine and fatigued Ti-6Al-4V/SCS-6 composites [9,20], where the reverse ISS was found to be lower than the forward ISS.

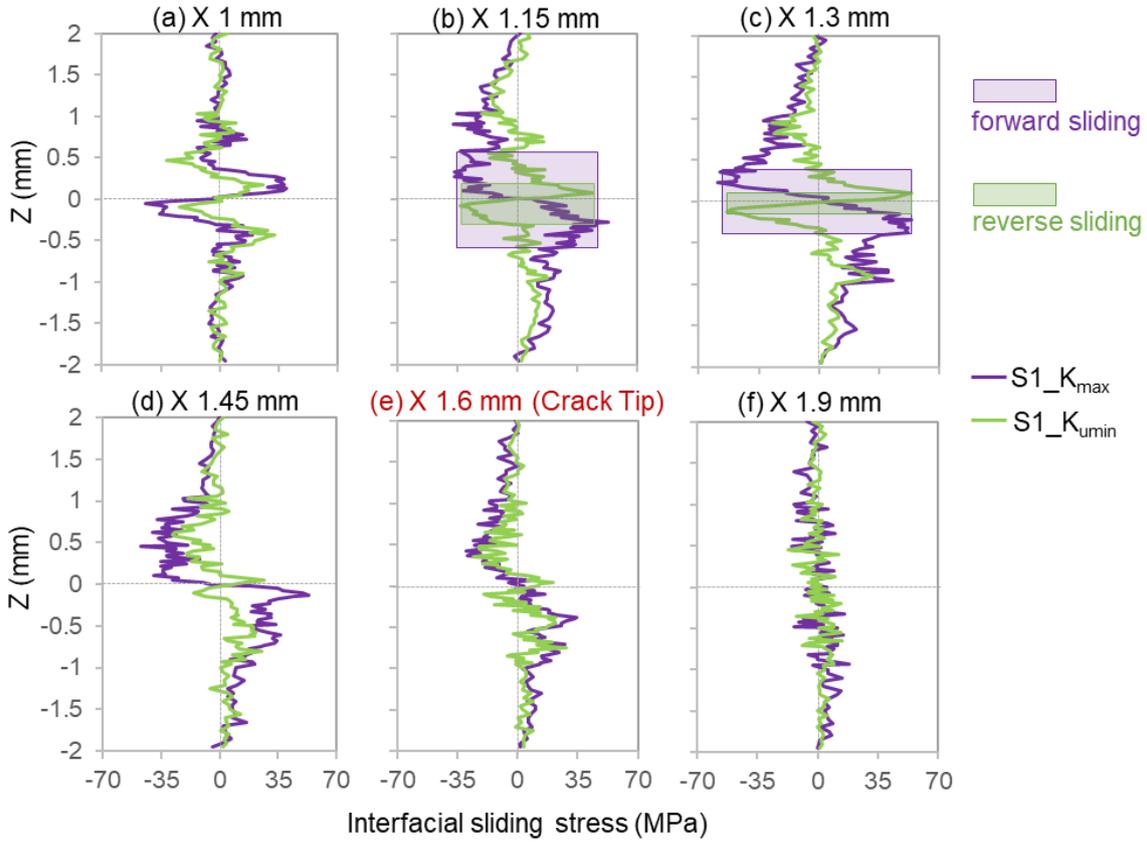

Fig. 8. Variation in interfacial sliding stress (inferred from the corresponding fibre strains in Fig. 7) along the fibres at different X positions for S1 in the peak loaded and unloaded states at 350 °C. The ISS profiles are shown at X positions of (a-d) 1, 1.15, 1.3, 1.45 (in the crack wake), (e) 1.6 (at the crack tip) and (f) 1.9 mm (ahead of crack tip) at $K_{max}$ and $K_{umin}$.

### 3.2 Behaviour in the rapid growth regime

For the higher $\Delta K_{ini}$ case (S2), during the *in situ* loading, a burst of crack growth accompanied with fibre fractures occurred as the stress approached $K_{max}$. The crack morphology and the stress redistribution micro-mechanisms in the presence of rapid crack growth and fibre fractures are discussed in this section.

#### 3.2.1 Crack morphology and fibre fractures in 3D

Fig. 9 shows a magnified virtual XZ section in S2 before ($K_{mean}$) and after ($K_{max}$) rapid crack growth. It is worth noting that although many fibres in the fatigue crack wake remained intact therefore bridging the crack during the stable growth observed during the fatigue pre-cracking, a few fibre fractures are evident in the CT scan at $K_{mean}$. A number of broken fibres can be seen in the longitudinal slice shown in Fig. 9(a) most of which lie very close to the crack plane (see yellow arrows). However, the relatively high fatigue $\Delta K_{ini}$ has also been sufficient to break one fibre (highlighted by the white arrow) distant from the crack plane (~350 μm). It is clear by comparing Fig. 9(a) and (b) that as $K_{max}$ was approached a burst of fibre fractures occurred leading to the rapid advance of the crack. Cotterill and Bowen [6] reported instantaneous rises in crack growth rate of up to two orders of magnitude (up to ~ $10^{-4}$ mm/cycle at 350 °C), associated with one or a group of fibre fractures, during fatigue crack propagation in Ti-15-3/SCS6 composite at elevated temperatures (200-500 °C). The rapid crack growth observed here is even more pronounced causing the crack to advance by 1 mm instantaneously. It is clear that the fibre fractures have enabled the fatigue crack formed during the stable-growth stage to open (see Fig. 9)., In contrast to the fibres that fractured during pre-fatigue which show fractures normal to the applied load, as seen in Fig. 9(b) many of the fibres that fractured in the rapid burst exhibit several wedge/disc-shaped fractures, which have the same morphological characteristics as those observed in the static axial tensile failure of UD Ti-6Al-4V/SCS6 SiC fibre composites at room temperature [34] as shown in the magnified inserts (i and iii) in Fig. 9.

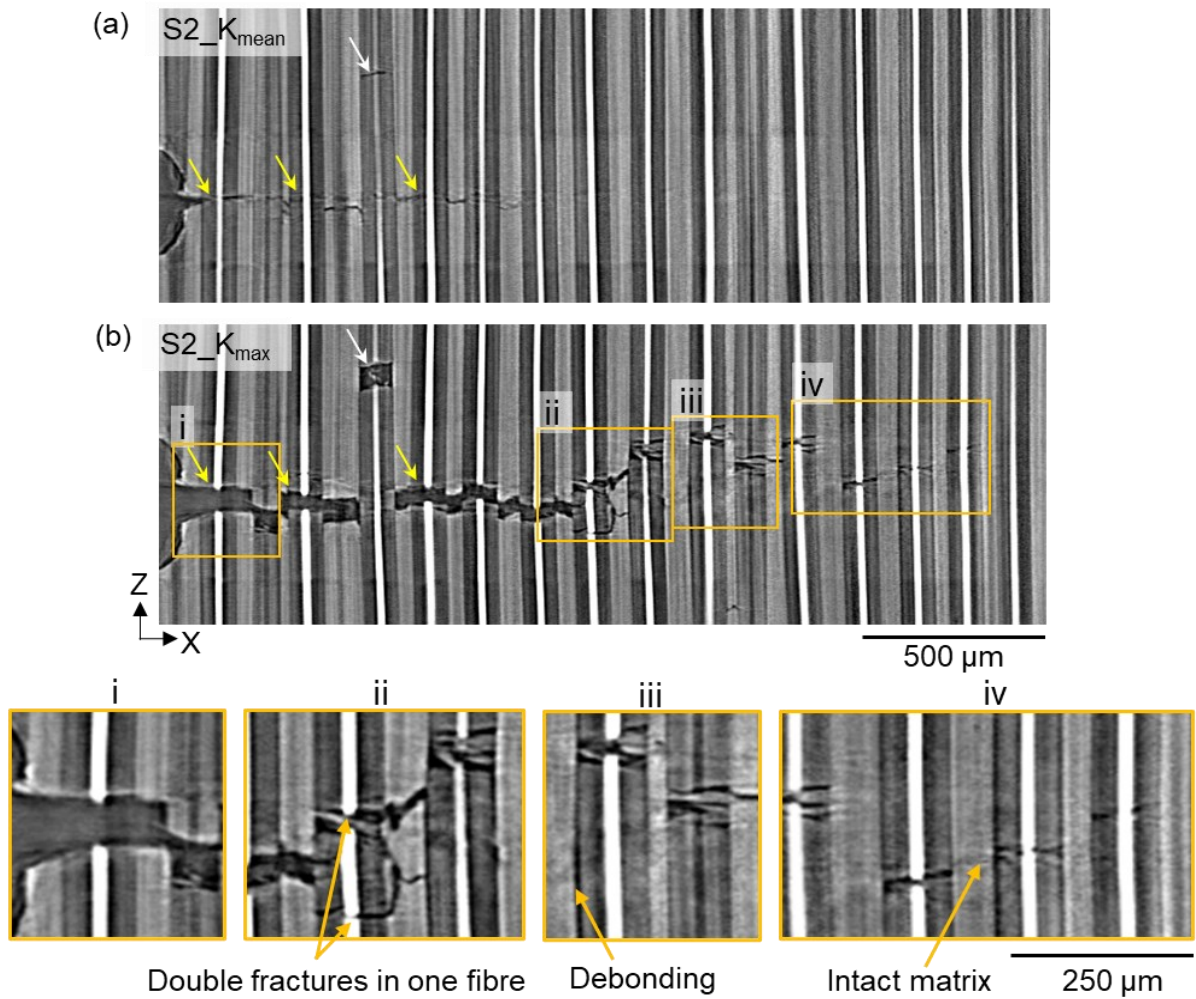

Fig. 9 X-ray CT virtual sections for S2 at 350 °C comparing damage a) before (at $K_{mean}$, 11.6 MPa√m) and b) after (at $K_{max}$, 19.0 MPa√m) a rapid burst of crack propagation. Yellow and white arrows indicate fibres that were damaged during the pre-fatigue. Inserts (i-iv) show magnified views of a range of different features marked in boxes in (b).

A stack of 201 XY sections centred at the notch-height (z = 0) in S2 were processed in MATLAB to obtain an image with the darkest pixel (with minimum grey value, which corresponds to crack/air) of all pixels in Z (i.e. having the same X and Y positions) plotted in a single XY slice image. (see Fig. 10) The projected XY sections show that the crack propagated predominantly in the fibre-reinforced region restrained somewhat by the tougher matrix-only cladding which may have also prevented catastrophic fracture of the entire specimen.

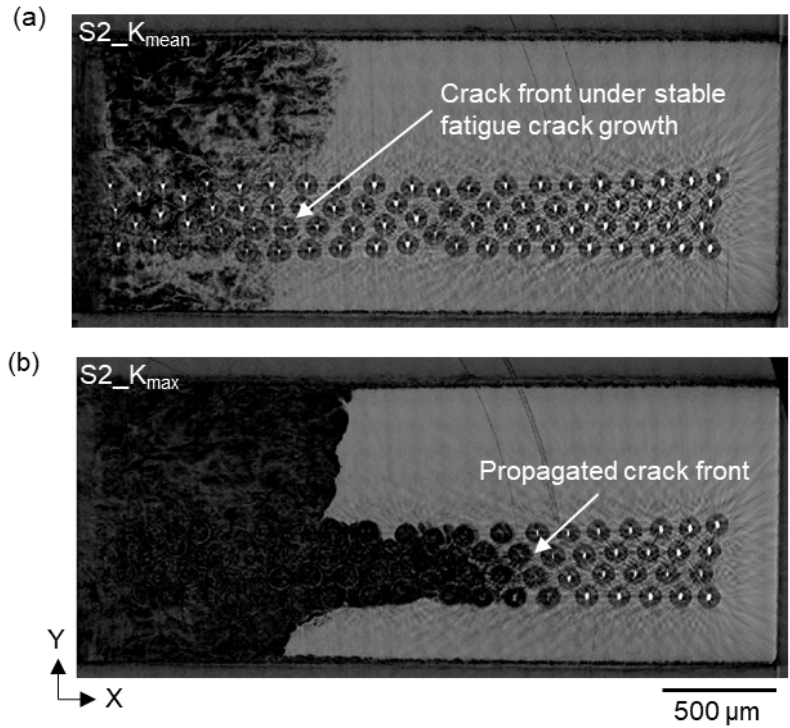

Fig. 10. X-ray CT virtual XY cross-sections of S2 at 350 °C comparing the fatigue crack a) at $K_{mean}$ and b) after rapid crack growth (at $K_{max}$). The black regions (corresponding to air) indicate where the crack has propagated.

Fig. 11 and Fig. 12 show the 3D morphology and topography of the crack before and after the period of rapid growth. It is evident that compared with the region corresponding to stable fatigue growth, the region of rapid growth is somewhat less tortuous. Furthermore, while the fibre fractures in the stable-growth zone are located close to the matrix crack plane, those that have fractured ahead of the original crack position are distributed over a much larger distance extending further away from the matrix crack plane, which could be because they formed ahead of the matrix crack. This is corroborated by the presence of fibre fractures ahead of the crack-tip position (see the 3D rendering in Fig. 11b), which is typical of the static tensile failure of UD Ti/SiC fibre composite [34]. This means there is less fibre pull-out in the stable fatigue-growth zone (see Fig. 13(a)) than in the rapid-growth zone (see Fig. 13(b)) on the final fracture surface. In Fig. 11(c), the fibre fractures have been colour-coded based on their Y positions into four plies. Although the fibre arrangement in the four plies of fibres is slightly different, we have not observed any evidence of the fibre mis-arrangement influencing the crack path.

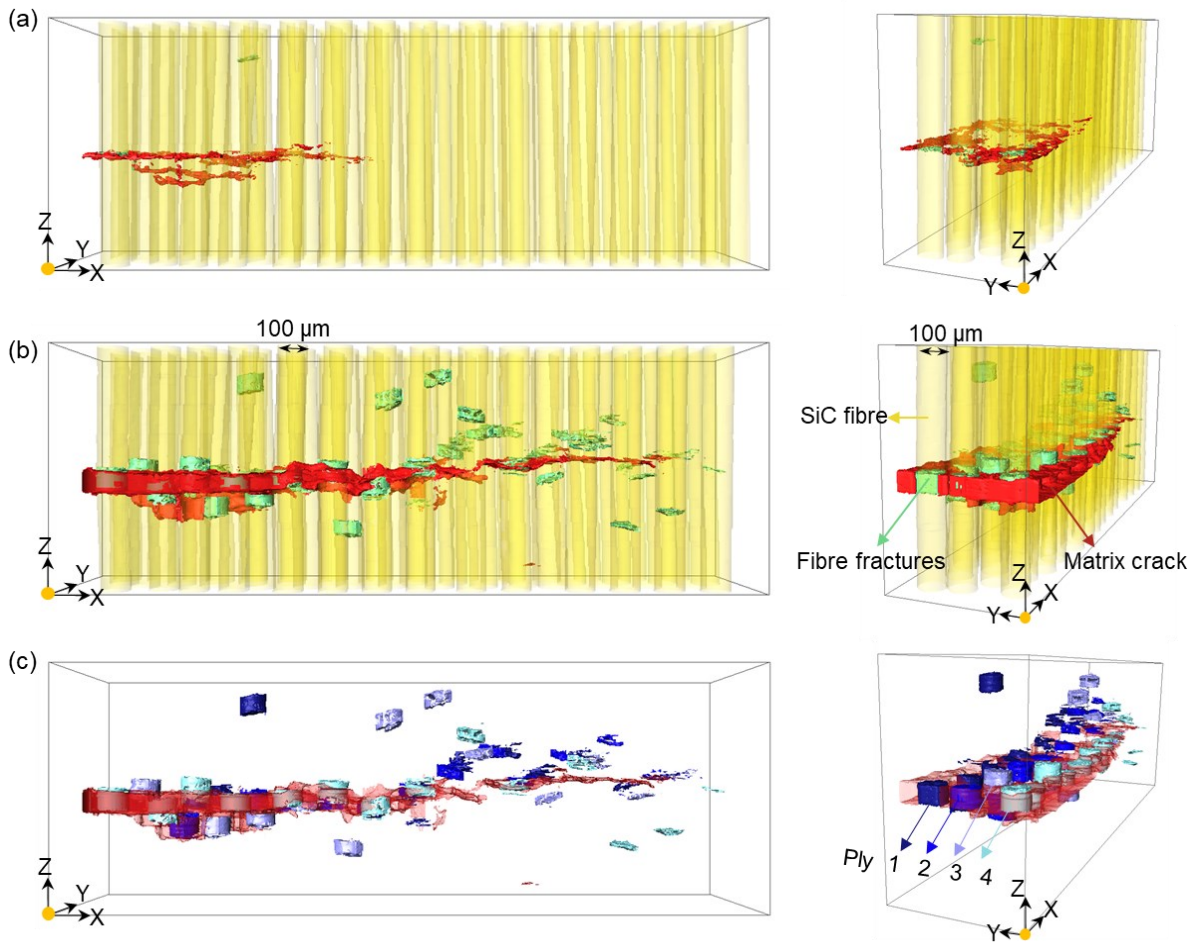

Fig. 11. 3D morphology of crack in S2 before (at $K_{mean}$) and after (at $K_{max}$) the period of rapid crack growth, showing the spatial distribution of the matrix crack and fibre fractures. Extracted and 3D volume rendered X-ray CT image of the crack (a) before and (b) after the period of rapid crack growth (the front on views (LHS) and perspective views (RHS) of the matrix crack (red), SiC fibres (yellow) and fibre fractures (green)), and (c) the fibre fractures are grouped in four plies (represented by different shades of blue) in S2 at $K_{max}$ with the matrix crack rendered semi-transparent.

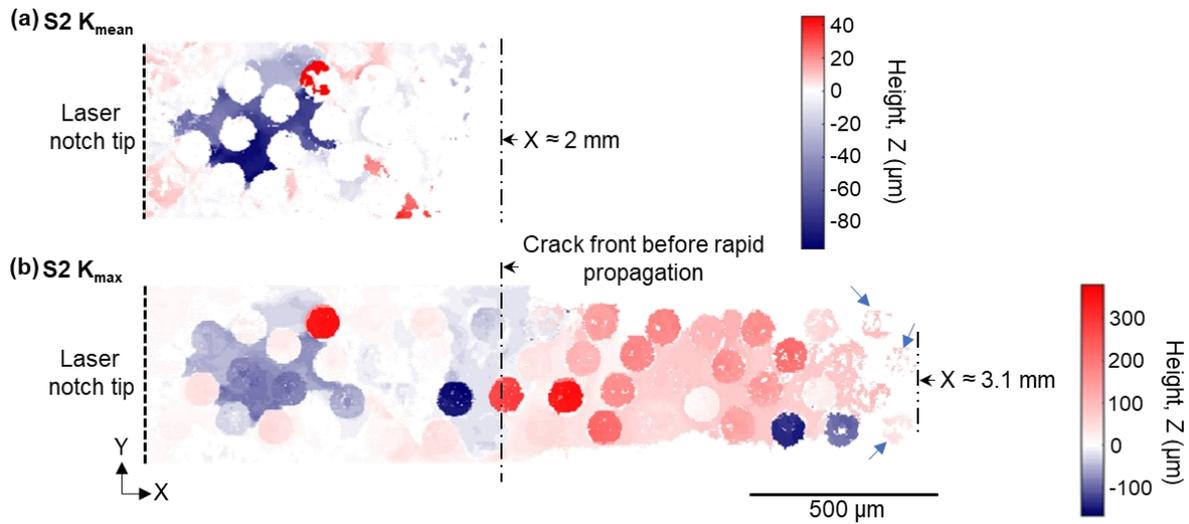

Fig. 12. The map of crack height showing crack deflection and the height of fibre fractures (a) before and (b) after the period of rapid crack growth. The height of the crack plane above and below z = 0 (the height of notch tip) was calculated based on the X-ray CT segmentation of S2 at $K_{mean}$ and $K_{max}$.

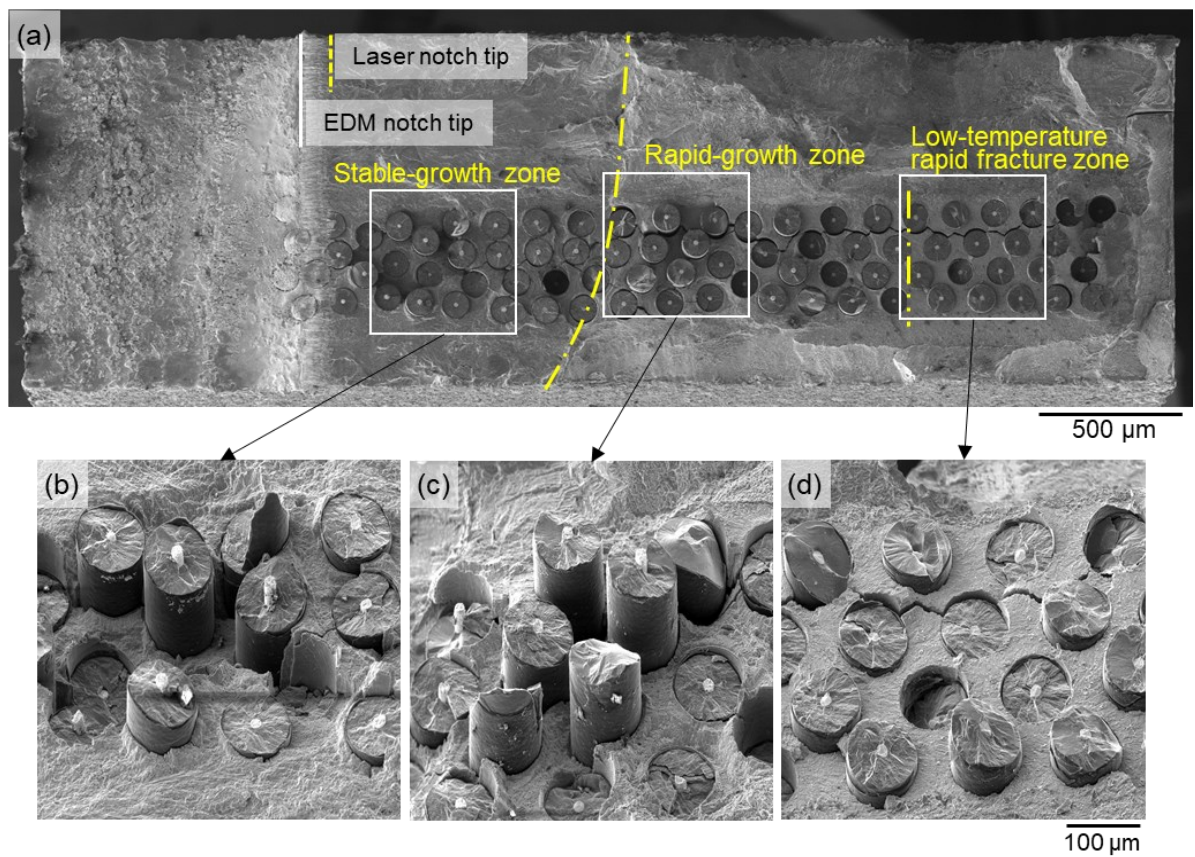

Fig. 13. The SEM images of the fracture surface of S2. (a) An overall micrograph and magnified views, showing (b) the stable fatigue crack growth zone and (c) the rapid crack growth zone developed at 350 °C, and (d) the zone that corresponds to the fast fracture in liquid nitrogen.

*3.2.2  Stress redistribution and interfacial sliding after rapid crack growth*

Fig. 14(a-b) shows the axial fibre elastic strain maps for S2 before and after rapid crack growth. Comparing the fibre strain map of S2 at $K_{mean}$ (11.6 MPa√m) (see Fig. 14(a)) with that of S1 at $K_{max}$ (16.5 MPa√m) after the same number of cycles (see Fig. 7(a)), where the damage states correspond to fatigue crack development during stable growth, the fibre strain map of S2 shows a larger wedge-shaped region characteristic of load redistribution as a result of the higher applied load. The effect of the rapid burst of fibre fractures is evident both in terms of the changes in fibre strain field between $K_{mean}$ and $K_{max}$ (see Fig. 14(a-b)) but also by comparison of the fields at $K_{max}$ for S1 (see Fig. 7(a)) and S2 (see Fig. 14(b)). The fibre strains in the vicinity of the crack plane (z = 0 mm) have fallen to zero (white in Fig. 14(b)), confirming that the fractures evident in Fig. 11 (extending up to x ≈ 3.1 mm) mean that the fibres were not supporting any load across the crack. The positions of fibre fractures (rendered brown) are generally found to match the regions with zero fibre strain (in white). Note that the white (low stress) region is larger in the rapid-growth zone than in the stable-growth zone, which can be attributed to the relatively larger scatter in the location of the fibre fractures away from the crack plane in this zone. Ahead of the crack tip, the maximum elastic fibre strain is about 0.50% which is similar to the maximum fibre strain measured in the bridging fibres in S1 (0.52%). In addition, the fibre strain falls sharply from tensile strain into compressive strain across the remaining un-cracked ligament (3<x<4 mm), which may be due to the bending component induced by crack opening on the remaining (~0.9 mm long) un-cracked ligament (shown at $K_{max}$ and $K_{umin}$ in Fig. 14(b)). The fact that the strain in the remaining intact fibres is not very high explains why the rapid crack growth and the large number of fibre fractures did not lead to rupture of the sample. This is due to the load borne by the un-cracked cladding in the wake of the propagated crack seen in Fig. 10(b).

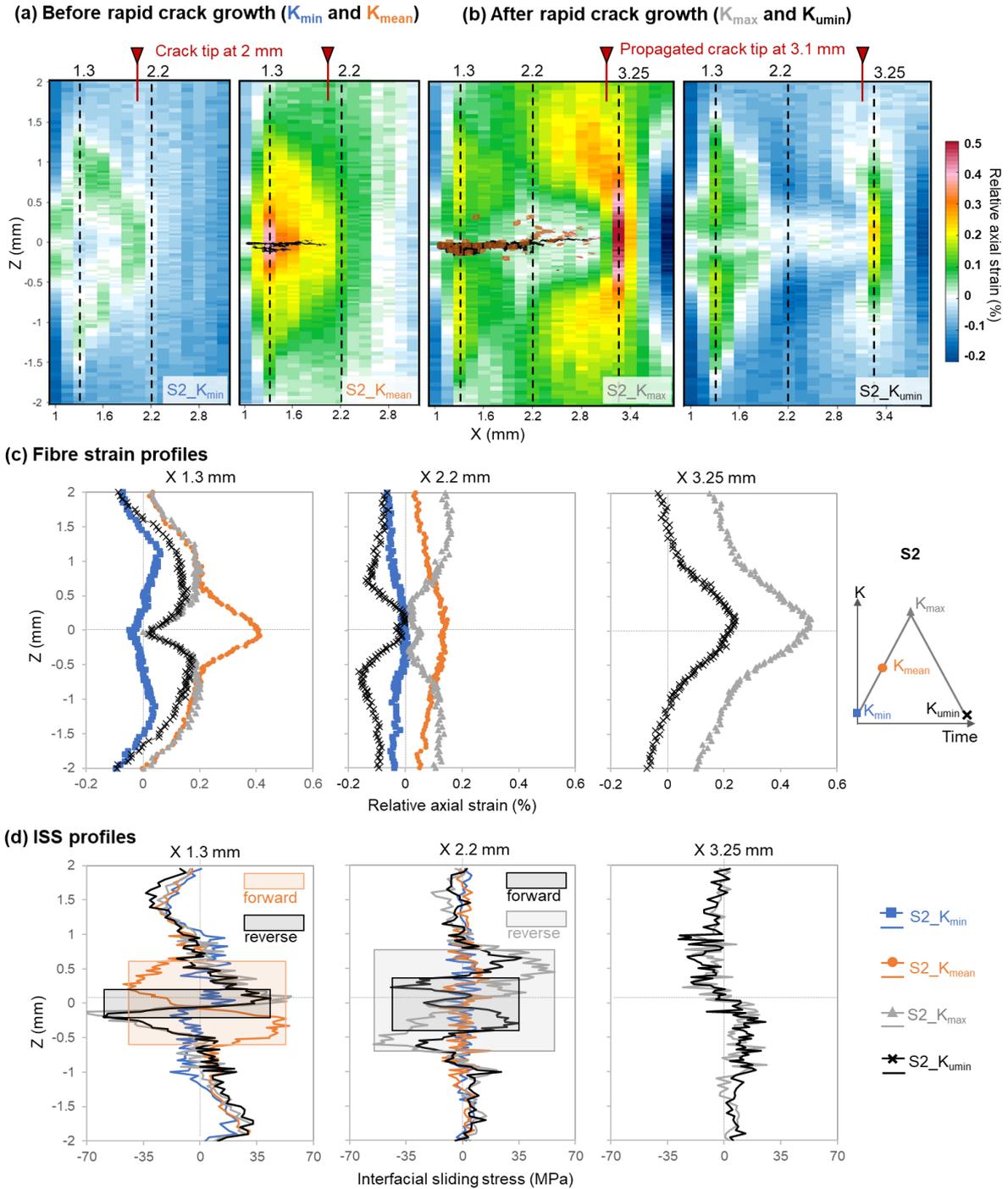

Fig. 14. Variation in relative axial fibre elastic strain distribution and interfacial sliding stress ahead of the notch in S2 at 350 °C. The fibre strain maps over a loading cycle, (a) before rapid crack growth at $K_{min}$ and $K_{mean}$, and (b) after rapid crack growth at $K_{max}$ and $K_{umin}$, with the extracted matrix crack (in black) and the fibre fractures (in brown) overlaid for $K_{mean}$ and $K_{max}$. The dashed lines correspond to the X positions for (c) and (d). The crack tip is at x ≈ 2 mm at $K_{min}$ and $K_{mean}$, whereas at x ≈ 3.1 mm at $K_{max}$ and $K_{umin}$. (c) The fibre elastic strain profiles at X positions of 1.3, 2.2 and 3.25 mm. (d) The interfacial sliding stress (inferred from the corresponding fibre strains) along the fibres at the X positions in (c).

Further, the influence of load redistribution mechanisms upon fibre fracture is investigated via the line profiles of fibre strains and the ISS profiles (see Fig. 14(c-d)). Three characteristic regions have been analysed: 1) the region where the bridging fibres have fractured (x = 1.3 mm), 2) the region originally ahead of the crack containing dispersed fractured fibres (x = 2.2 mm) and 3) ahead of the final crack tip position (x = 3.25 mm). At x = 1.3 mm, the peak in the strain profile at z = 0 mm at $K_{mean}$ clearly indicates the tensile load borne by the fibres bridging the crack. This falls to zero after rapid crack growth due to the resulting fibre factures. The ISS profile at $K_{mean}$ is indicative of forward sliding (fibre pull out) at a frictional shear stress of around 50 MPa as for S2 and a sliding length of around 600 μm either side of the crack plane. After fracture the fibres retract over a distance of around 200 μm which, as for the fibres at x = 1 mm in S1, then remains retracted as the load is cycled from $K_{max}$ to $K_{umin}$. Prior to the burst of rapid crack growth the crack tip is positioned at around 2 mm. Consequently the stress at x = 2.2 mm at $K_{mean}$ is considerably below that at the crack tip (see Fig. 14(a)) and there is no evidence of frictional sliding along the fibres located there in common with the fibres just ahead of the crack for the low stress amplitude sample (S1). After fibre fracture at $K_{max}$ the fibre stresses (at z = 0 mm) fall to zero and the fibres retract significantly (sliding over a length of ~700 μm) with a frictional sliding stress around 55 MPa. Then upon unloading the fibres are pushed out somewhat over a sliding length around 400 μm. For the fibres ahead of the propagated crack tip (x = 3.25 mm), the stress profile is very similar to that ahead of the crack for S1. There is no evidence of interfacial sliding because the shear stresses here at $K_{max}$ are very low.

## 4    Conclusions

In this study, the damage accumulation and load partitioning mechanisms of TC17 titanium alloy/SiC fibre composites at a temperature representative of likely in-service aeroengine environments have been investigated by simultaneously visualising the 3D damage morphology (by X-ray CT) and strain mapping (by XRD) at 350 °C in the stable and rapid growth regimes.

In the low stress amplitude case, stable fatigue crack growth is accompanied by a fatigue crack that propagated without breaking the SiC fibres. In this case the tensile load supported by the bridging fibres shield the fatigue crack, reducing the effective stress intensity factor at the crack tip. The matrix crack is observed to have deflected (by 50-100 µm in height) when bypassing bridging fibres. The fibres bridging the crack are observed to support stresses up to ~2100 MPa at $K_{max}$ in the crack wake. This loading of the fibres is achieved through the interfacial shear stress. It is found that in this case the fibres slide out of the matrix at $K_{max}$ by as much as 600 µm in the crack wake and the closer the fibres are to the crack tip the shorter the sliding distance. The frictional sliding stress is around 55 MPa.

In the higher stress amplitude fatigue case, even in the stable growth regime some broken fibres are observed, mostly lying near to the crack plane but a few further from it. Loading to the $K_{max}$ for this test led to a rapid growth of the crack accompanied by a burst of fibre factures. Many fibres were fractured with those lying ahead of the original crack position breaking at a greater distance from the crack plane compared to those that were originally bridging the crack giving rise to higher levels of fibre pull-out. Given that there were some cracked fibres ahead of the final crack position, this suggests that some of these fibres may have broken ahead of the advancing crack. The fibres that fractured in this region tend to have a more complex crack morphology than those that failed during stable crack growth. In this case the fractured fibres retract into the matrix upon fracture over a sliding length of around 700 µm at a frictional sliding stress around 55 MPa.

The maximum frictional sliding stress is ~55 MPa in both regimes in the fatigued TC17/SiC fibre composites at 350 °C. This is higher than the 20-40 MPa measured at 300 °C in fatigued Ti-6Al-4V/SCS-6 as reported in [20]. Overall the results show the importance of maintaining the fatigue stress amplitude below the level capable of breaking the bridging fibres and that even though stable growth may be possible, such a scenario runs the risk of a sudden burst of fibre fractures and catastrophic failure which proceeds under different mechanisms and different interactions with the composite structure.

**Acknowledgements**

This work was supported by Beijing Institute for Aeronautical Materials. We acknowledge Diamond Light Source for time on Beamline I12 under Proposal EE20226. The authors would like to thank Thomas Connolley and Kaz Wanelik at Beamline I12 and Luke Rollings at The University of Manchester for their help and support throughout the beamtime. This work was also supported by the Henry Moseley X-ray Imaging Facility (funded through EPSRC grants EP/F007906/1, EP/F001452/1, EP/I02249X, EP/M010619/1, EP/F028431/1 and EP/M022498/1) within the Henry Royce Institute for Advanced Materials (funded through EPSRC grants EP/R00661X/1, EP/S019367/1, EP/P025021/1 and EP/P025498/1). PJW acknowledges the support of an ERC Advanced Grant (CORREL-CT) grant agreement number 695638.